\documentclass[%
 reprint,
 amsmath,amssymb,
 aps,
]{revtex4-2}

\usepackage[utf8]{inputenc}
\usepackage{graphicx}
\usepackage{dcolumn}
\usepackage{amsmath}
\usepackage{bm}
\usepackage{amsmath}
\usepackage{physics}
\usepackage{verbatim}
\usepackage{color}
\usepackage{tabularx}
\usepackage{xcolor}

\newcommand\COMMENTED[1] {}

\begin{document}

\title{Algorithm for branching and population control in correlated sampling}
\preprint{APS/123-QED}

\author{Siyuan Chen}
\thanks{These two authors contributed equally}
\affiliation{Department  of  Physics,  College  of  William  \&  Mary,  Williamsburg,  Virginia  23187, USA}
\author{Yiqi Yang}
\thanks{These two authors contributed equally}
\affiliation{Department  of  Physics,  College  of  William  \&  Mary,  Williamsburg,  Virginia  23187, USA}
\author{Miguel Morales}
\affiliation{Center for Computational Quantum Physics, Flatiron Institute, 162 5th Avenue, New York, New York, USA}
\author{Shiwei Zhang}
\affiliation{Center for Computational Quantum Physics, Flatiron Institute, 162 5th Avenue, New York, New York, USA}

\begin{abstract}
Correlated sampling has wide-ranging applications in Monte Carlo calculations. When branching random walks are involved, as commonly found in many algorithms in 
quantum physics and electronic structure, population control is typically not applied with correlated sampling due to technical difficulties. This hinders the stability and efficiency of correlated sampling. In this work, we study schemes for allowing birth/death in correlated sampling and propose an algorithm for population control. The algorithm can be realized in several 
variants depending on the application. One variant is a static method that creates a reference run and allows other correlated calculations to be added \textit{a posteriori}. Another optimizes the population control for a set of correlated, concurrent runs dynamically. These approaches are tested in different applications in quantum systems, including both the Hubbard model and electronic structure calculations in real materials.
\end{abstract}
\maketitle

\section{introduction}
Monte Carlo methods\cite{Kalos2008} are widely used in engineering\cite{Jain_Book1991}, physics\cite{Zhang2019,Foulkes_RMP2001,Spanier_Book1969,Howell_AHT1969,Bertini_AC1976}, reliability theory\cite{Ayyub_Book2003}, queuing theory\cite{Rubinstein_Book1986}, finance\cite{Mcleish_Book2005}, etc. In many applications, Monte Carlo methods are used to solve the underlying differential equations or integral equations of the processes or systems in a stochastic way. This is often achieved by random walks that are typically constructed by Markov chains, where the transition probability is designed to reflect the characteristic of the system. 

In a broad class of algorithms, a random walker gains a weight that fluctuates as the random walk proceeds, leading to an exponentially increasing variance\cite{Hetherington_PRA1984}. To solve this problem, the random walk is designed to be carried out by multiple walkers simultaneously with branching\cite{Hetherington_PRA1984,Kalos2008,Trivedi_PRB1990,Buonaura_PRB1998}. Branching random walks duplicate the walkers with large weights and eliminates walkers with small weights under certain probabilities. In addition, to avoid the unbounded fluctuation of the total population, population control is generally applied\cite{Hetherington_PRA1984,Buonaura_PRB1998,Nguyen_CPC2014}. During the random walk, samples are collected and the weighted average of all the samples gives a Monte Carlo representation of the target quantity.
Branching random walk algorithms are widely applied in the study of ground-state\cite{Foulkes_RMP2001,Zhang_PRB1997,Zhang_PRL2003} (and even finite-temperature\cite{finiteT-RW-PhysRevLett.83.2777}) properties of interacting quantum many-body systems. 

In quantum physics or chemistry, the difference between two closely related systems is often of great interest and importance. Examples include binding energies and gaps, redox potential, reactions, etc. In addition, other observables, such as forces in molecules and solids\cite{Motta_JCP2018, SC_2023_ForceStress}, order parameters in lattice models\cite{Qin_PRX2020}, etc., can be obtained by computing energy derivatives based on finite difference. Quantities computed from Monte Carlo methods naturally contain statistical errors. When taking their differences, the statistical errors are propagated into the target quantity. If the differences are small, the signal-to-noise ratio would be low. 

Correlated sampling\cite{Kalos2008} is generally applied in the Monte Carlo calculation of the energy differences and gradients to reduce the statistical noise. A considerable amount of work exists to develop and apply correlated sampling to various forms of quantum Monte Carlo methods, including 
variational Monte Carlo calculation of potential energy curves \cite{cs_Umrigar_IJQC1989} and particle-hole excitations\cite{cs_Kwon_PRB1994}, Green's function Monte Carlo (GFMC) and diffusion Monte Carlo (DMC) calculation of potential energy curves and bond lengths\cite{cs_Filippi_PRB2000}, $H_3$ cation\cite{cs_Traynor_CPL1988}, the dipole moment of LiH\cite{cs_Wells_CPL_1985}, and forces as well as polarizabilities\cite{cs_Vrbik_JCP1992}, and phaseless auxiliary field quantum Monte Carlo (Ph-AFQMC) calculation of bond dissociation energies, ionization potentials, and electron affinities~\cite{Shee2017ChemicalTA}. 

Combining  branching random walks and correlated sampling is  not straightforward. The difficulty originates from branching de-correlating the random walks in different systems. Even when the systems are very close, branching can eventually cause the random walks to deviate and the correlations between them to then quickly deteriorate. To prolong correlation, typically correlated sampling calculations forgo branching and population control \cite{Shee2017ChemicalTA}, instead simply keeping the weights of walkers. This exacerbates the asymptotic instability of the calculation. To our knowledge population control algorithms have not been discussed in correlated sampling calculations, nor has there been a systematic study of how it can affect the efficiency, accuracy, and stability of 
the calculation. 

In this paper, we propose an algorithm for branching and population control in correlated sampling, which ensures stability and thus prolongs the correlation time. 
Our correlated sampling schemes with branching and population control are then tested in ground-state AFQMC calculations, in both the Hubbard model
and a periodic solid system. The statistical fluctuation is found to be significantly reduced compared with that of the runs without correlated sampling, and the correlations between the runs are sustained for 
significantly longer 
compared with correlated sampling without population control. 
Although  our population control algorithms 
are implemented and studied within AFQMC,
they apply to any correlated sampling calculations 
that involve branching random walks, as further discussed below. 

The rest of the paper is structured as follows. In Sec. \ref{sec:motivation}, we provide a brief overview of correlated sampling as well as branching and population control. This is followed in Sec. \ref{sec:population_control} by a description of our population control algorithms, together with discussions on the metrics to quantify the performance of correlated sampling. The test results in the Hubbard model and real materials are shown in Sec. \ref{sec:results}. Then we conclude in Sec. \ref{sec:conclusion}.

\section{Motivation}\label{sec:motivation}
The expectation value of a quantity $A(x)$ measured from a probability distribution of $x$ is expressed by an integral over the probability distribution function (PDF) $p(x)$:
\begin{equation}\label{eq:expectation}
    \left< A \right> = \frac{\int p(x) A(x) dx}{\int p(x) dx},
\end{equation}
where $p(x)$ can be, for example, the partition function $\mathcal{Z}$ in statistical physics, or the probability density $|\Psi|^2$ for a quantum state. The variance of $A$ describes the spread of $A$ for different $x$ values: 
\begin{equation}
    \mathrm{var}(A) = \frac{\int p(x) \qty(A(x)-\left< A \right>)^2 dx}{\int p(x) dx}.
\end{equation}
If the integral in Eq.(\ref{eq:expectation}) is in high-dimensional space, one typically uses the Monte Carlo method to evaluate it. 

Monte Carlo samples form a stochastic representation of the PDF $p(x)\rightarrow\sum_i w_i \delta(x-x_i)$. The Monte Carlo estimation of quantity $A$ is then given by:
\begin{equation}
    A_{\rm MC}=\frac{\sum_{n=1}^N w_n\, A(x_n)}{\sum_{i=n}^N w_n},
    \label{eq:MCest}
\end{equation}
which approaches the expectation value of $A$ as the number of samples $N$ increases: $\abs{\left< A \right>-A_{\rm MC}}\propto1/\sqrt{N}$~\cite{Krauth_Book2006,Zhang2019} according to the law of large numbers. 

\begin{figure}
    \includegraphics[width=0.5\textwidth]{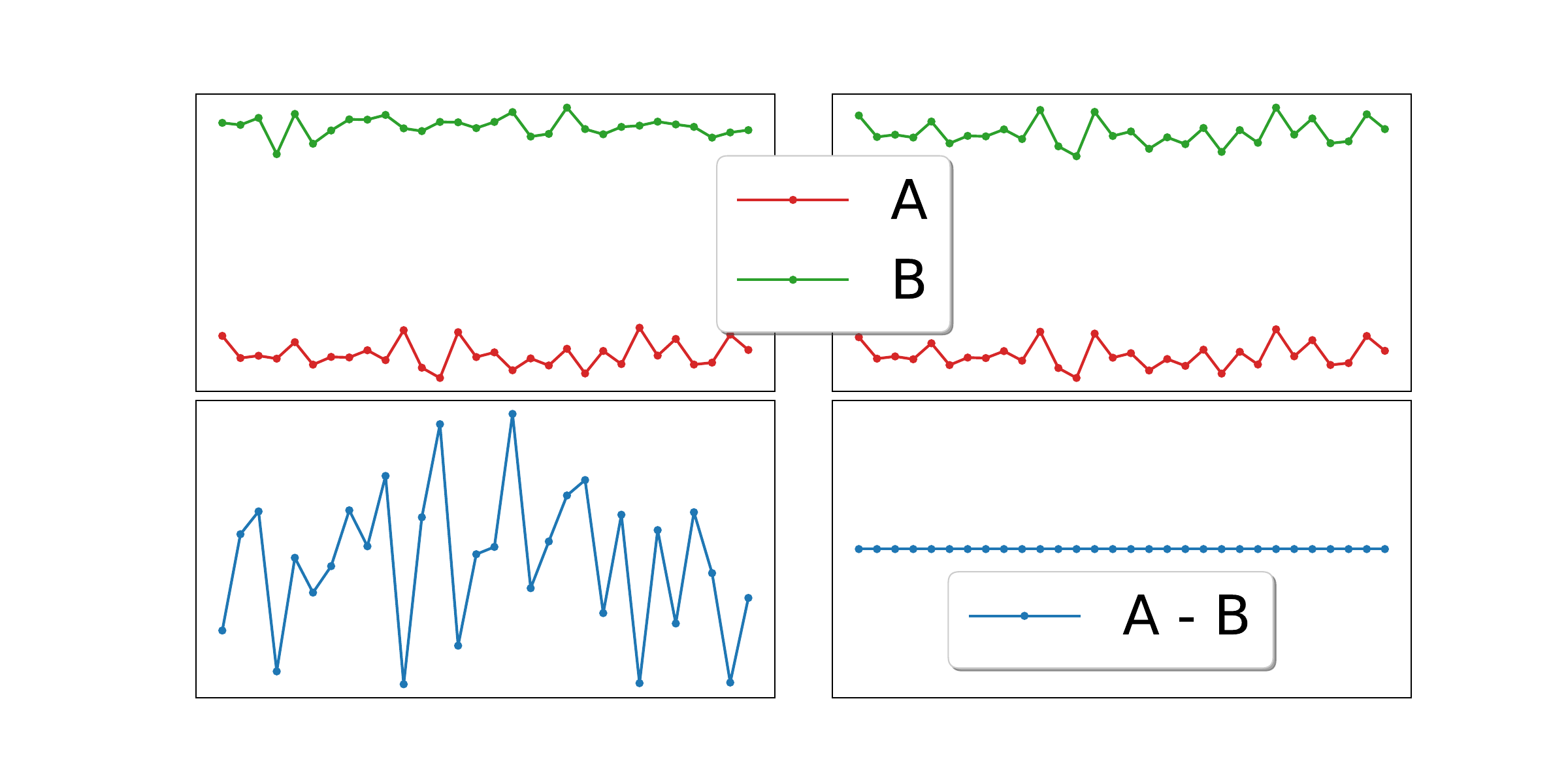}
    \caption{Schematic illustration of  
    the effect of  
    correlated sampling in computing the differences. The left panels show uncorrelated calculations while the right panels show correlated sampling. 
    Horizontal axes give Monte Carlo time, i.e., index of the branching random walk steps. The top panels give ``snapshot'' measurements in two separate 
    calculations for A and B respectively, while the bottom panels show the 
    computed difference.}
    \label{fig:SOC}
\end{figure}

Sometimes we are interested in the difference between two quantities, $A$ and $B$.
When computing the difference $A-B$ via Monte Carlo method, its variance $\mathrm{var}\qty(A-B)$ can be expressed as follows~\cite{Kalos2008,Motta_WIRES2018}:
\begin{equation}\label{motivation}
    \mathrm{var}\qty(A-B)=\mathrm{var}\qty(A)+\mathrm{var}\qty(B)-2\mathrm{Cov}\qty(A,B). 
\end{equation}
Here,
\begin{equation}
    \mathrm{Cov}\qty(A,B) = \left<AB\right> - \left<A\right>\left<B\right>
\end{equation}
denotes the covariance of the two quantities. 
If the two quantities are independently sampled, in other words their covariance is zero, the variance of the difference would be the sum of the variance of $A$ and $B$, as illustrated in the left column of Fig. \ref{fig:SOC}.
However, if one can correlate the sampling of the two quantities, for example by using the same random number stream, to maximize the covariance, the statistical error of their difference will be minimized. The ideal case occurs when correlation is perfect, $2\mathrm{Cov}(A,B)=\mathrm{var}\qty(A)+\mathrm{var}\qty(B)$, and $A-B$ is a constant so that the difference is obtained with no statistical fluctuation, as shown in the right panel of Fig.~\ref{fig:SOC}. This motivates the idea of correlated sampling.

\begin{figure}
    \includegraphics[width=0.5\textwidth]{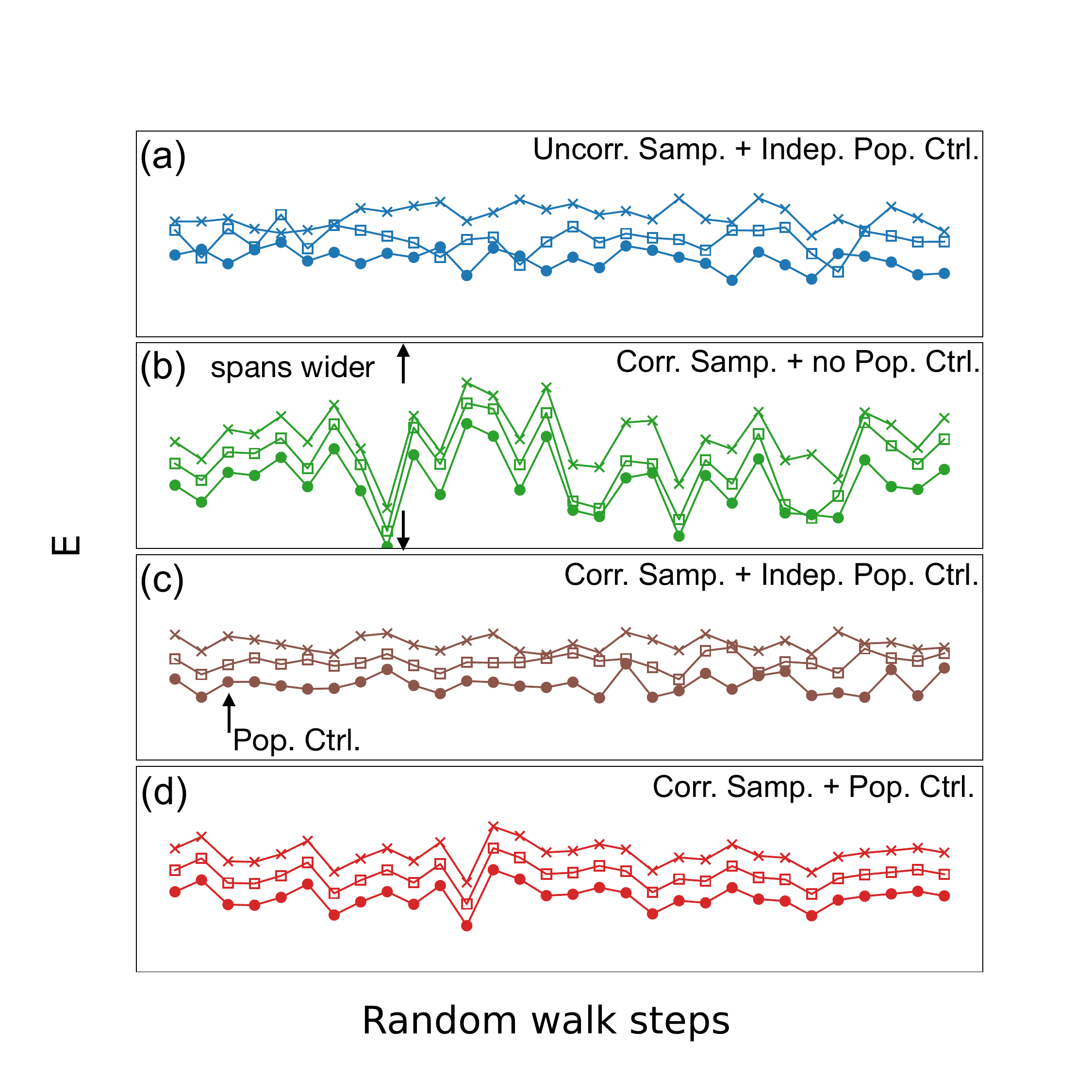}
    \caption{Schematic illustration of the conflict between correlated sampling and branching and population control. 
    (a) Uncorrelated calculations of three closely related systems incur
    independent noise and hence large fluctuations in estimating the differences. 
    (b) In correlated sampling without population control, runs remain correlated 
    as random walk proceeds, but large fluctuations in weights make 
    the statistical error grow rapidly.
    (c) In correlated sampling with independent population control, separate branching can cause the runs to de-correlate quickly.
    (d) Algorithm for branching and population control in correlated sampling aims to maximize efficiency and prolong correlation.
}
    \label{fig:SOI}
\end{figure}

In random walks that involve branching and population control, for instance, the AFQMC method\cite{Zhang_PRL2003}, the implementation of correlated sampling is not straightforward. This is due to an inherent conflict between correlated sampling and branching random walks, as illustrated in Fig.~\ref{fig:SOI}. Panel (b) shows the energy fluctuations for three correlated runs, without population control. Early on in the calculations (small number of random walk steps), the correlation is strong and the runs lead to much reduced fluctuations in the estimated differences. This mode has been very useful in quantum chemistry calculations \cite{Shee2017ChemicalTA,Shee_JCP2023}. As time goes on, however, the weights will fluctuate a lot more in the absence of branching and population control, causing larger statistical noise in each measured results from Eq.~(\ref{eq:MCest}), hence also larger noise in the differences. Doing individual and separate branching and population control in each run will break down the correlation rapidly, as shown in panel (c). Since the systems being correlated are not identical, their random walks branch with different rates. As soon as population control leads to different branching decisions (e.g., extra walker or elimination of a walker in one run), a correlation between the different runs will be lost. The idea of the present paper is to 
incorporate branching and population control in a correlated manner between the different runs (panel (d)), which allows the calculations to stay well correlated for a much longer time.

\section{Branching and population control in correlated sampling}\label{sec:population_control}

\begin{figure*}
    \includegraphics[width=0.8\textwidth]{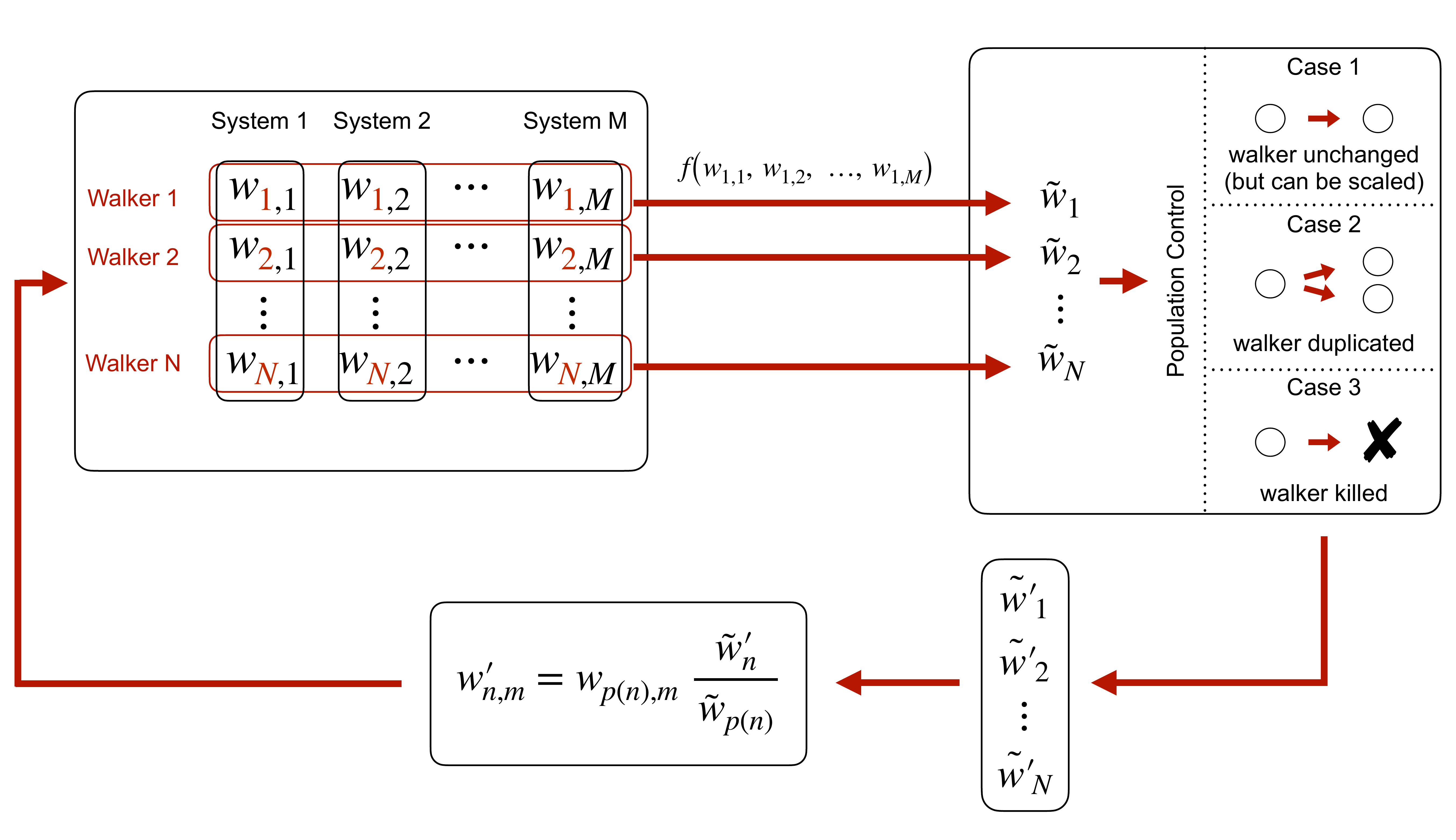}
    \caption{Flowchart of the branching and population control algorithm in correlated sampling. Walker weights prior (subsequent) to population control are denoted by $w_{p(n),m}$ ($w'_{n,m}$), where $p(n)$ is the index of the parent walker of the $n$th walker from the population control decision, with $M$ correlated systems/runs and $N$ walkers for each system. The reference weights, $\{ \tilde w_{p(n)}\}$, are obtained via $f$ and fed into population control, which gives output decisions represented by a new set of reference weights, $\{ \tilde w_n'\}$. The individual weights are then updated accordingly.
    }
    \label{fig:AFC}
\end{figure*}

In this section, we introduce our algorithm for branching and population control in correlated sampling. The algorithm is described in a flowchart in Fig.~\ref{fig:AFC}. We will assume that $M$ systems (runs) are correlated, each with a population size $N$. The weights of run $m$ are given by $\{ w_{n,m} \}$ with walker index $n=1,\cdots,N$. Walkers from different runs with index $n$ are correlated. As the walkers are propagated and branched, this set of correlated walkers must follow 
the same branching path to delay de-correlation.

To implement this, a reference weight is derived from each set of correlated walkers through some user-defined function:
\begin{equation}\label{eq:f-def}
    \tilde w_n \equiv f(w_{n,1},w_{n,2},\cdots,w_{n,M})\,. 
\end{equation}
A universal choice of $f$ is taken and applied to all $n$ but, as we further discuss below, the form of the function is flexible. For instance, $f$ could be the $\max$ function so as to take the maximum weight in each correlated walker set, or the average function, or it could take the weight of a specific system (fixed $m$, but see below), etc. The reference weights $\{ \tilde w_n \}$ are then fed into a ``normal'' branching and population control routine, i.e., one that is conventionally used in the branching random walk without correlated sampling. This outputs a set of branching and population control decisions on the reference weights, in the form of a new set of weights, $\{ \tilde w'_{n} \}$. We then update the actual walker weights in each run by keeping constant the ratio between the new and old weights:
\begin{equation}\label{eq:scale}
    w'_{n,m}= w_{p(n),m}\,
     \frac{\tilde{w}'_{n}}{\tilde{w}_{p(n)}},
\end{equation}
where 
$p(n)$
is the index of the parent walker for $n$ from the population control decision. 

Various realizations exist for the branching and population control algorithm, in part depending on the choice of $f$, which can lead to different implementations that emphasize different 
aspects of the computation or programming and can have relative advantages and disadvantages:
\begin{enumerate}
\item 
Carry out all the correlated runs concurrently, where the reference walker weights are derived from each correlated walker set on the fly. In this way, the branching and population control are optimized dynamically, and we refer to this as the ``dynamic'' realization. All the choices for 
$f$ mentioned above are possible within this scheme.
\item
Conduct a single Monte Carlo run as the reference run and record the reference weights, which store the branching and population control decisions made therein. Other correlated runs (``child runs'') are then added freely afterward, following the reference weight and branching decisions in the reference run. We refer to this as the ``static'' realization. In this scheme, the function $f$ must be based on a single
run index $m$ which is the reference run.
\item 
Combining (1) and (2), one can record the branching and population control decisions from a dynamic run, then add subsequent runs where the decisions are applied. We refer to this as a ``semi-dynamic'' realization.
\end{enumerate}
We summarize the characteristics of the three methods in Table~\ref{tab:popCtrlCompare}.
\begin{table}[b]
\caption{\label{tab:popCtrlCompare}
Characteristics of the three implementations of the branching and population control algorithm in correlated sampling.
}
\begin{ruledtabular}
\begin{tabular}{cccc}
 &add runs&synchronous run&\# reference runs\\
\hline
Dynamic& N & Y & $>1$ \\
Static& Y & N & 1 \\
Semi dynamic & Y & Y& $>1$ \\ 
\end{tabular}
\end{ruledtabular}
\end{table}

In the static realization in (2), the function $f$ has limited choices, since we only have knowledge of the walker weights in the reference system. We also need to consider that walkers with zero (or very small) weights need to be carried through and not population controlled, because other unknown runs might have nonzero (or much larger) weights. Our choice for the reference weight for the static realization of population control is
\begin{equation}\label{static_ref_weight}
    \tilde{w}_n =
\begin{cases}
    w_{n,1},&   w_{n,1}>w_\mathrm{th}\\\
    w_\mathrm{safe}\,.              & \text{otherwise}
\end{cases}
\end{equation}
where $m=1$ labels the reference run, $w_\mathrm{th}$ is a small positive number, and $w_\mathrm{safe}$ is a positive number that ensures a walker with $w_{n,1} = w_\mathrm{th}$ or smaller will yield exactly one copy in the population control. Alternatively, for general population control algorithms, one can simply use $\tilde{w}_n=w_{n,1}$, and manually prevent branching in all runs for this walker when $w_{n,1} \leq w_\mathrm{th}$. 

The goal of a correlated branching and population control algorithm is to prolong the time during which the systems remain well correlated. Even with branching and population control, the correlation will degrade eventually for each set of correlated walkers and for the entire systems. Small differences in walker weights in a set of correlated walkers accumulate over time to become large differences, and the probability of a rare event (e.g., one walker in a correlated run is killed while others retain a significant weight) will also increase.

We monitor the level of correlation with a number of metrics. For example, we can take the standard deviation of the walker weight for each correlated walker group and ``normalize'' it with respect to its mean as a measure of the relative weight fluctuation. The average across the entire population
\begin{equation}\label{eq:wt-fluc-Q}
    Q=\sum_{n=1}^N\sqrt{\frac{M\sum_m w_{n,m}^2}{(\sum_m w_{n,m})^2}-1}\bigg/N\,
\end{equation}
then gives a simple global indicator. A $Q$ value larger than a few, for instance, would indicate a significant ``dispersion'' of the weight among the correlated runs. This way we could examine the quality of the correlation on the fly (as a function of random walk steps), and stop the calculation when it reaches a certain level. The walker weights are the only quantities needed in this process.

\begin{figure}[b]
\includegraphics[width=0.45\textwidth]{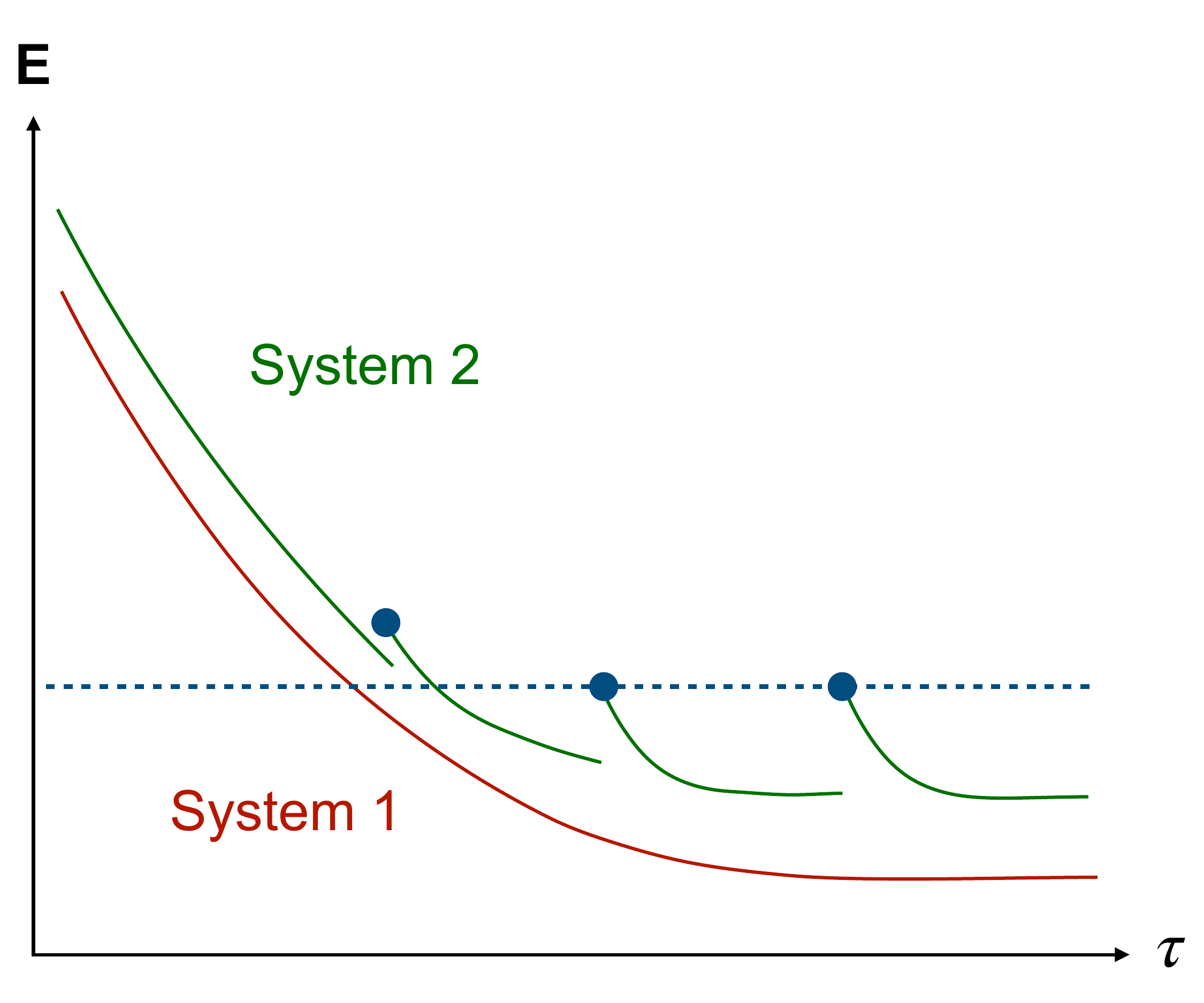}
\caption{Schematic illustration 
of the reset method used in dynamic correlated sampling. 
The computed energies are shown for two systems as a function of 
random walk time steps (projection time $\tau$ in Ph-AFQMC) starting 
from equilibration. System 1 is chosen (arbitrarily) as the reference.
At reset, the population of System 2 is replaced 
by that of System 1, which causes the jumps in the computed expectation value in System 2. The jump magnitude reaches a constant when both systems have 
equilibrated. Extra steps are needed for re-equilibrating System 2 
after every reset.
}
\label{fig:SOR}
\end{figure}

To maximize utilization of the well-correlated period, we minimize the equilibration time by starting correlated sampling with an equilibrated walker set from one of the correlated systems. This is called ``preliminary equilibration scheme'' and was proposed by Shee et al. previously~\cite{Shee2017ChemicalTA}. To improve statistics and reduce error bars, it is typically more efficient to repeat the entire procedure multiple times with different random seeds. If the correlation between the correlated runs remains significant (e.g., as indicated by a small $Q$ metric value) beyond the auto-correlation time of the measurements in the individual runs, then we can make multiple measurement blocks to collect more data. 

In the dynamic realization, instead of the ``preliminary equilibration scheme'', we reset the concurrent runs periodically to repeat the sampling, achieving an effective re-start at every reset which accomplishes the same goal. Specifically, a reset period is chosen before the calculation; once the propagation imaginary time reaches this specified reset period, we reset the run by copying the walkers of one of the correlated systems (we call it the reference system and it can be chosen arbitrarily) to other systems. In Fig.~\ref{fig:SOR} we show a schematic of the reset method for dynamic branching and population control in correlated sampling, using Ph-AFQMC as an example. Note that in both schemes, a reweighting is needed, as the underlying importance functions are typically different between the two systems. 
(In Ph-AFQMC, this is reflected by the different trial wave functions, $|\Psi_{T1}\rangle$ and $|\Psi_{T2}\rangle$ for Systems 1 and 2, respectively.) When the population of System 2 is replaced by the surrogate population from System 1, the probability density is modified, hence the jumps in the computed expectations in the illustration. The dynamic reset method provides a natural way to 
continue the correlated sampling runs to improve statistics, more in the spirit of regular branching random walks. 

\section{Results}\label{sec:results}

In this section, we show test results in two very different classes of problems. We use the 
AFQMC method \cite{Zhang_PRB1997,Zhang_PRL2003} applied to the ground state of the Hubbard model and to solid silicon. They provide a spectrum of tests in real-world problems which are sufficiently challenging and of strong current research interest in quantum physics. Both static and dynamic realizations of the population control algorithm
will be tested. The different types of quantum Monte Carlo algorithms (a lattice problem with local interactions in the former and a continuum problem with long range interactions in the latter) allow us to do so under different, general conditions. 
Rather than 
obtaining new physical results, our focus here is on testing, illustrating, and studying the population control algorithm and correlated sampling. The physical quantities we will compute are simple and well understood, so that we can benchmark the results straightforwardly.

The implementation details are very different for the two classes of problems we consider here, as well as the behaviors of the algorithms. We refer the readers to the literature for further discussions. For understanding the tests below, it suffices to recall that the AFQMC method \cite{Zhang2019,Motta_WIRES2018,Shi_JCP2021} projects out the ground state of a Hamiltonian $\hat H$ by applying $e^{-\tau \hat H}$, to yield 
\begin{equation}\label{AFQMC-psi}
\ket{\psi(\tau)}\propto\sum_{n} w_n^{(\tau)} \ket{\phi_n^{(\tau)}}/\braket{\psi_T}{\phi_n^{(\tau)}}\,,
\end{equation}
where $\tau$ denotes projection time, and $\ket{\psi(\tau)} \xrightarrow{\tau\rightarrow \infty} |\psi_G\rangle$, the  desired ground state. The projection is realized by a branching random walk whose time step is represented by $\tau$, in which $n$ is the index of the random walker and $w_n$ is the weight. The random walker $\ket{\phi_n}$ lives in a space of Slater determinants determined by the details of the problem, and $\braket{\psi_T}{\phi_n}$ is an importance function defined by a known trial wave function  $\ket{\psi_T}$. The random walkers are propagated in the manifold of Slater determinants via a set of random auxiliary fields, ${\mathbf x}$: $B({\mathbf x}) |\phi^{(\tau)}\rangle \rightarrow |\phi^{(\tau')}\rangle $ (with $\tau'=\tau+\Delta \tau$ being the next step in the random walk). In this framework it is convenient and straightforward to think of  correlated sampling as correlating the multi-dimensional auxiliary-fields ${\mathbf x}$. (See Ref.~\cite{Zhang_PRB1997} for a way to cast diffusion Monte Carlo \cite{Foulkes_RMP2001} in this framework.) We will correlate multiple runs for different but related Hamiltonians, $\hat H_m$,  with different pre-defined $\ket{\psi_{T,m}}$, producing $\ket{\psi_m(\tau)}$. It is important to note that, in the context of the population control algorithm, the weights $w_{n,m}^{(\tau)}$ are the only objects we will need to deal with. 

\subsection{Hubbard Model}\label{Hubbard}

We study the ground state of the Hubbard model on a square lattice~\cite{Hubbard_PRSLA1963}, 
one of the fundamental models in quantum many-body physics:
\begin{equation}
    \hat{H} = -t\sum_{\langle ij\rangle\sigma}\hat{c}_{i\sigma}^\dagger\hat{c}_{j\sigma} + U\sum_i\hat{n}_{i\uparrow}\hat{n}_{i\downarrow}\,,
    \label{eq:H-Hub}
\end{equation} 
where $t$ is the hopping parameter, $U$ the on-site interaction, $\hat{c}_{i\sigma}^\dagger$ ($\hat{c}_{j\sigma}$) creates (annihilates) an electron with spin $\sigma$ at site $i$ ($j$), and the $\langle...\rangle$ refers to the nearest-neighbor hopping. We will focus on computing the double occupancy
\begin{equation}\label{eq:D-def-Hub}
    D\equiv \langle \psi_G|\sum_i\hat{n}_{i\uparrow}\hat{n}_{i\downarrow}|\psi_G\rangle\,,
\end{equation}
which is directly proportional to the interaction energy and which provides an important measure to the nature of electron correlation. 
(Note that the physical ``double occupancy'' on 
each lattice site is given by $D$ 
divided by the number of lattice sites.)
Using Hellman-Feynman theorem
\begin{equation}\label{eq:Hellman-Feynman}
    D=\langle \psi_G|\frac{d\hat{H}}{dU}|\psi_G\rangle=\frac{dE_G}{dU}\,,
\end{equation}
where the derivative on the right-hand side can be evaluated by finite difference
\begin{equation}
    \frac{dE_G}{dU}\bigg\vert_{U}
    \doteq \frac{E_G(U+\delta U)-E_G(U-\delta U)}{2\,\delta U}\,. 
    \label{eq:FD-D}
\end{equation}
Thus the double occupancy $D$ for $\hat H$ can be computed by the ground-state energy difference between two Hamiltonians $\hat H_1$ and $\hat H_2$, with $U$ in Eq.(\ref{eq:H-Hub}) replaced by $U_1=U+\delta U$ and $U_2=U-\delta U$, respectively. 

\begin{figure}[b]
\includegraphics[width=0.45\textwidth]{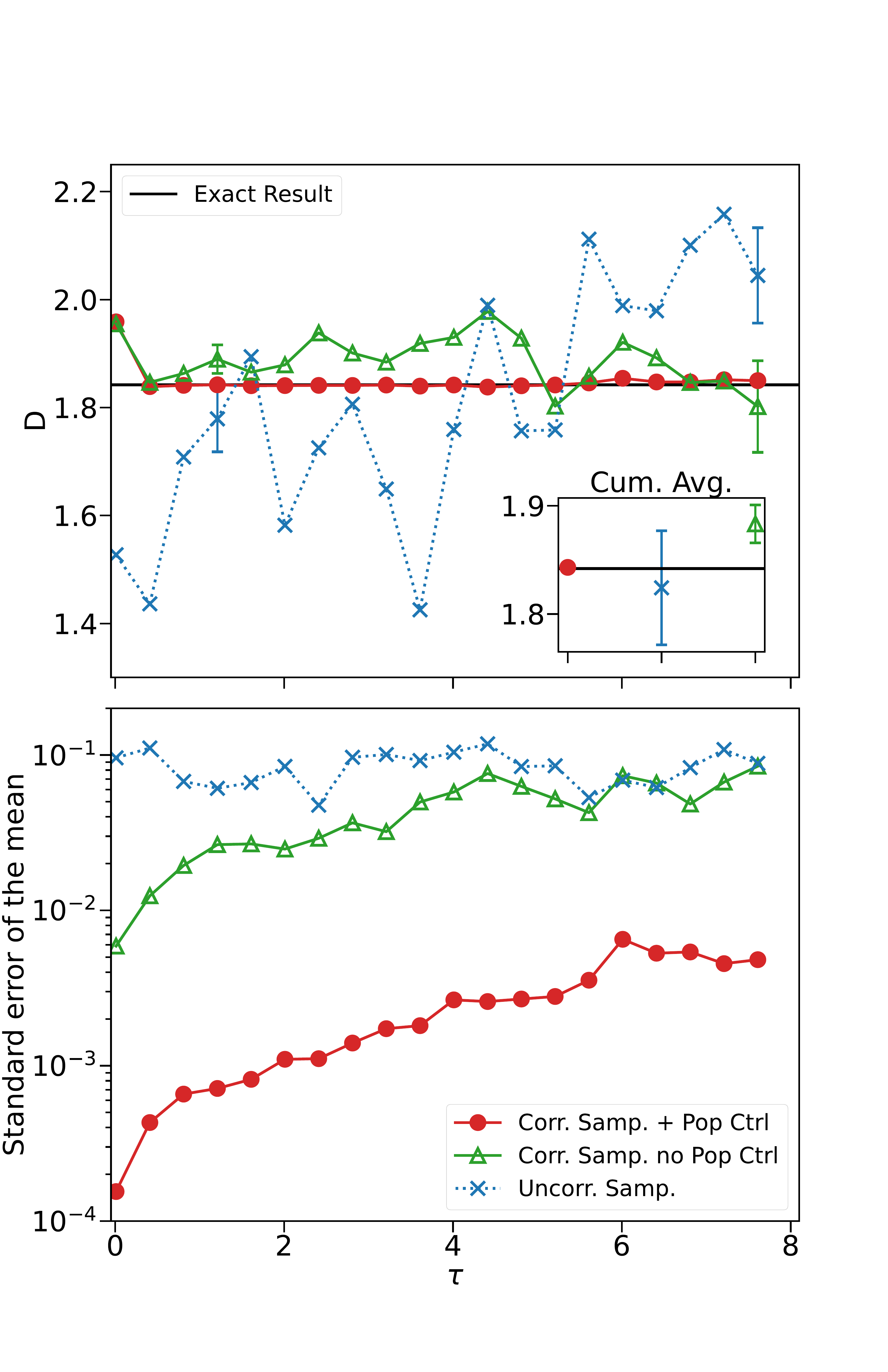}
\caption{Double occupancy results $D$ as a function of projection time (random walk steps) obtained from uncorrelated, correlated sampling with no population control, and correlated sampling with population control calculations using the finite difference shown in Eq.~(\ref{eq:FD-D}). The corresponding standard error for each $\tau$ is shown as an error bar in the upper panel at $\tau=1.21$ and $7.61$ only for the sake of clarity, but is plotted for each data point in the lower panel. These error bars are estimated from 40 resets (or 40 repeated runs for the uncorrelated calculations). The cumulatively averaged results are shown in the inset. (Accumulations are taken after discarding an equilibration time $\tau=1.2$.) The horizontal black solid line appearing in both the main plot and the inset in the upper panel is the exact answer obtained from exact diagonalization. The system is  $L_x=L_y=4$, $U/t=4$, at half-filling with periodic boundary conditions, with $\delta U=0.01$. Note the log scale.
}
\label{fig:CPD}
\end{figure}

We test the dynamic realization for branching and population control in correlated sampling to compute the double occupancy $D$. We first study a small system of lattice size $4\cross4$, $U/t=4.0$, at half-filling with periodic boundary conditions along both $x$ and $y$ directions. The trial wavefunctions were generated from the generalized Hartree-Fock method with effective onsite interaction $U_{\rm eff}$~\cite{Qin_PRB2016,Qin_PRB2016_sc}. 
The benchmark results are shown in Fig.~\ref{fig:CPD}. Here each random walk step corresponds to an increment of $\tau$ of $\Delta\tau=0.01$. Results at the same imaginary time relative to the reset points are averaged to estimate the statistical error. We can see that results are all consistent with the exact answer. In the uncorrelated run, the statistical fluctuation is large throughout the imaginary time propagation. While correlated sampling without population control shows a significantly reduced error bar at small $\tau$, this reduction deteriorates as $\tau$ increases, as expected. Population controlled correlated sampling is seen to significantly 
extend the correlation, reducing the statistical error by more than a factor of 10 throughout the entire convergence window in $\tau$. 

\begin{figure}[b]
\includegraphics[width=0.45\textwidth]{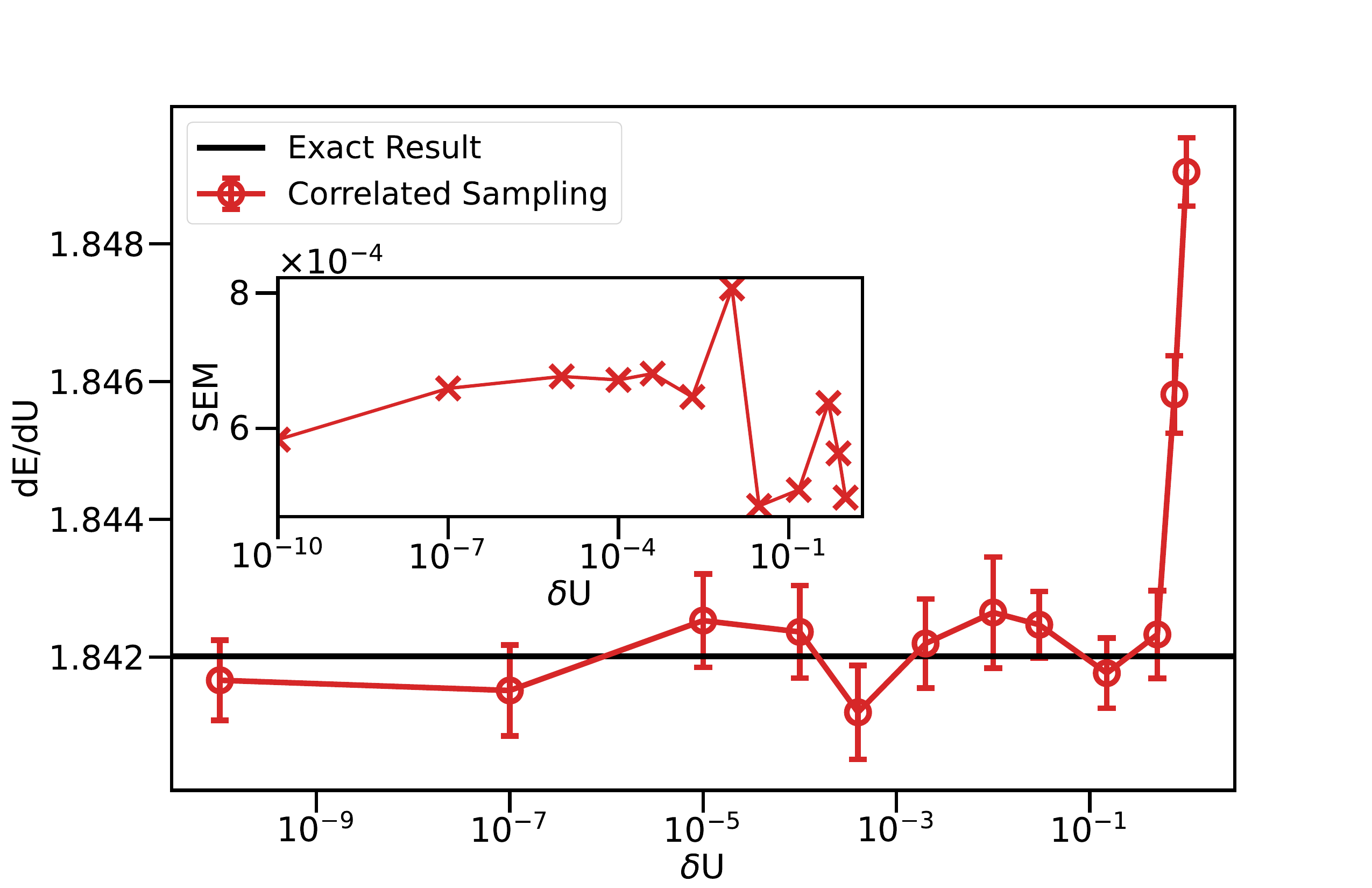}
\caption{Accuracy of correlated sampling calculations of the energy 
derivative versus finite-difference size. Results are shown for a range of $\delta U$ across many decades, for the same system as in Fig.~\ref{fig:CPD}. The main figure is the computed result and the inset shows the estimated error bar. Computational cost is the same for each calculation. Note the log scale on the x-axes. The exact result is obtained from exact diagonalization.
}
\label{fig:DVD}
\end{figure}

We next examine the behavior of correlated sampling versus the proximity of the correlated systems, which is specified by $\delta U$ in the present system. In uncorrelated runs, the statistical error grows in inverse proportion to $\delta U$, as given by Eq.~(\ref{eq:FD-D}).
The computed energy derivative $dE/dU$ with correlated sampling is shown in Fig.~\ref{fig:DVD}, for a range of $\delta U$ values spanning many decades. The results are surprisingly robust. Only at very large $\delta U$ values are systematic biases seen, indicating the breakdown of the finite difference formula in approximating the derivative. 
The statistical error is seen to be essentially independent of $\delta U$. This is highly advantageous, as very small values of $\delta U$ can be used to ensure that the finite difference yields an accurate estimate of the derivative, without increase in computational 
cost.

\begin{table}
\caption{\label{tab:testResults}
Double occupancy computed from calculations using correlated sampling with population control (``Corr.~Samp.'') and uncorrelated independent runs (``Uncorr.~Samp.''). Results are for 
the Hubbard model at half-filling with $U/t=8$, in periodic supercells of size $4\times 4$, $8\times 8$ and $16\times 16$. Note that the listed values are $D$ as defined in Eq.~(\ref{eq:D-def-Hub}), not per site. Statistical errors are in the last digit(s) as indicated in the parentheses.
}
\begin{ruledtabular}
\begin{tabular}{cccccc}
\textrm{L}&
\textrm{Corr. Samp.}&
\textrm{Uncorr. Samp.}&
\textrm{Efficiency Gain}\\
\colrule
$4\times 4$  & 0.857(2)  & 0.92(9) & $5.5\times10^3$\\
$8\times 8$  & 3.472(5) & 2.95(32) &$1.4\times10^4$\\
$16\times 16$  & 13.93(4) & 13.0(14) &$5.7\times10^3$\\
\end{tabular}
\end{ruledtabular}
\end{table}

The algorithm works equally well in larger system sizes, and major computational efficiency gain is seen in realistic system sizes in state-of-the-art calculations \cite{Qin_PRB2016,Xu_PRR2022}. In Table~\ref{tab:testResults} we show a simple comparison of correlated sampling with population control versus uncorrelated calculations of the double occupancy. For both types of calculations, $\delta U$ was fixed at 0.01. This is a reasonably safe choice to ensure that the finite difference approximation in Eq.~(\ref{eq:FD-D}) remains reliable. The systems have $U/t=8$, which are more strongly correlated than the previous example. In all the runs here, the error bars are estimated from 40 resets (or 40 repeat runs in uncorrelated calculations) after equilibration. We quantify the computational efficiency gain as
\begin{equation}
    {\left(\frac{\sigma_{\mathrm{uncorrelated}}}{\sigma_{\mathrm{correlated}}}\right)}^2\frac{T_{\mathrm{uncorrelated}}}{T_{\mathrm{correlated}}}\,,
\end{equation}
where $T$ is the total computational cost of a calculation (correlated or uncorrelated) and $\sigma$ is the corresponding statistical error bar. A smaller choice of  $\delta U$ would make the computational efficiency gain grow, inversely proportional to $\delta U^2$.

\begin{figure}[b]
\includegraphics[width=0.48\textwidth]{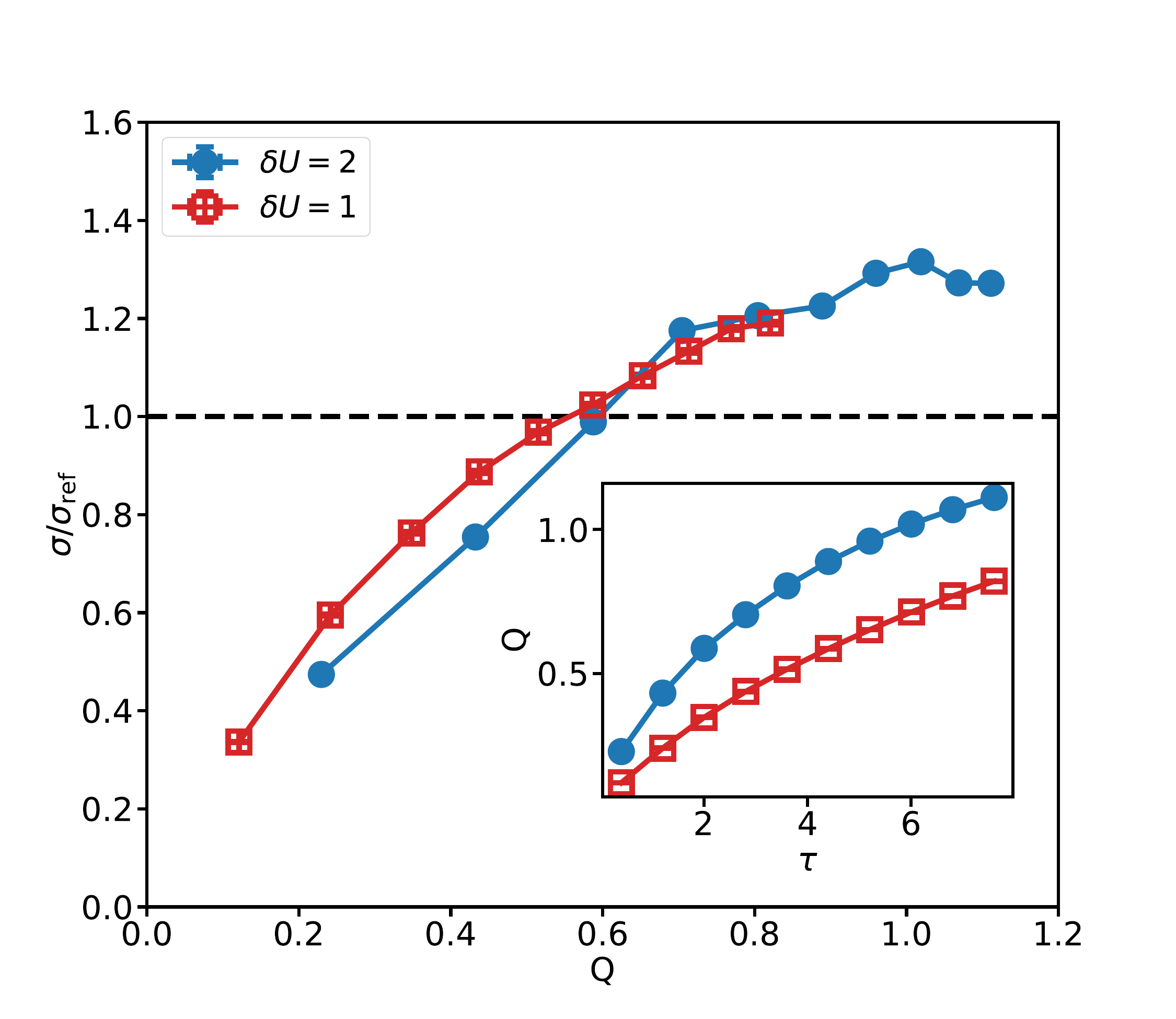}
\caption{Monitoring the level of correlation in 
correlated sampling.
The efficiency gain over uncorrelated calculations, indicated by the variance ratio $\sigma/\sigma_{\mathrm{ref}}$, is plotted in the main figure versus the metric $Q$
of Eq.~(\ref{eq:wt-fluc-Q}). Two examples with $\delta U=1$ and $\delta U=2$ are shown, for the same system as in Fig.~\ref{fig:CPD}. The horizontal dashed line marks $\sigma/\sigma_{\mathrm{ref}}=1$. The inset plots $Q$ versus  projection time $\tau$. All the curves were smoothed by averaging many neighboring points.
}
\label{fig:Q}
\end{figure}

In Fig.~\ref{fig:Q}, we illustrate 
how the index $Q$ defined in Eq.~(\ref{eq:wt-fluc-Q}) can help monitor  
the quality of the correlation. We quantify the superiority of the correlated sampling calculation over the uncorrelated calculation through the variance ratio, $\sigma/\sigma_{\mathrm{ref}}$, where $\sigma_{\mathrm{ref}}$ denotes the variance of the uncorrelated sampling calculation, which works as a reference. When this ratio exceeds 1, there is no longer any efficiency gain with correlated sampling. In order to see the crossover more clearly, we deliberately choose unnaturally large $\delta U$ values. 
As is shown in the main figure, $\sigma/\sigma_{\mathrm{ref}}$ increases with $Q$. In the inset, we see that $Q$ increases with projection time and, at the same $\tau$ value, $Q$ is larger for larger $\delta U$. 
These observations are all as expected from the nature of correlated sampling. The similar value of $Q$ where $\sigma/\sigma_{\mathrm{ref}}$ crosses 1 in the main figure indicates that $Q$ gives a reasonably generic metric for measuring the proximity of the correlated systems.

\subsection{Real materials: solid Si}

\begin{figure}[b]
\includegraphics[width=0.45\textwidth]{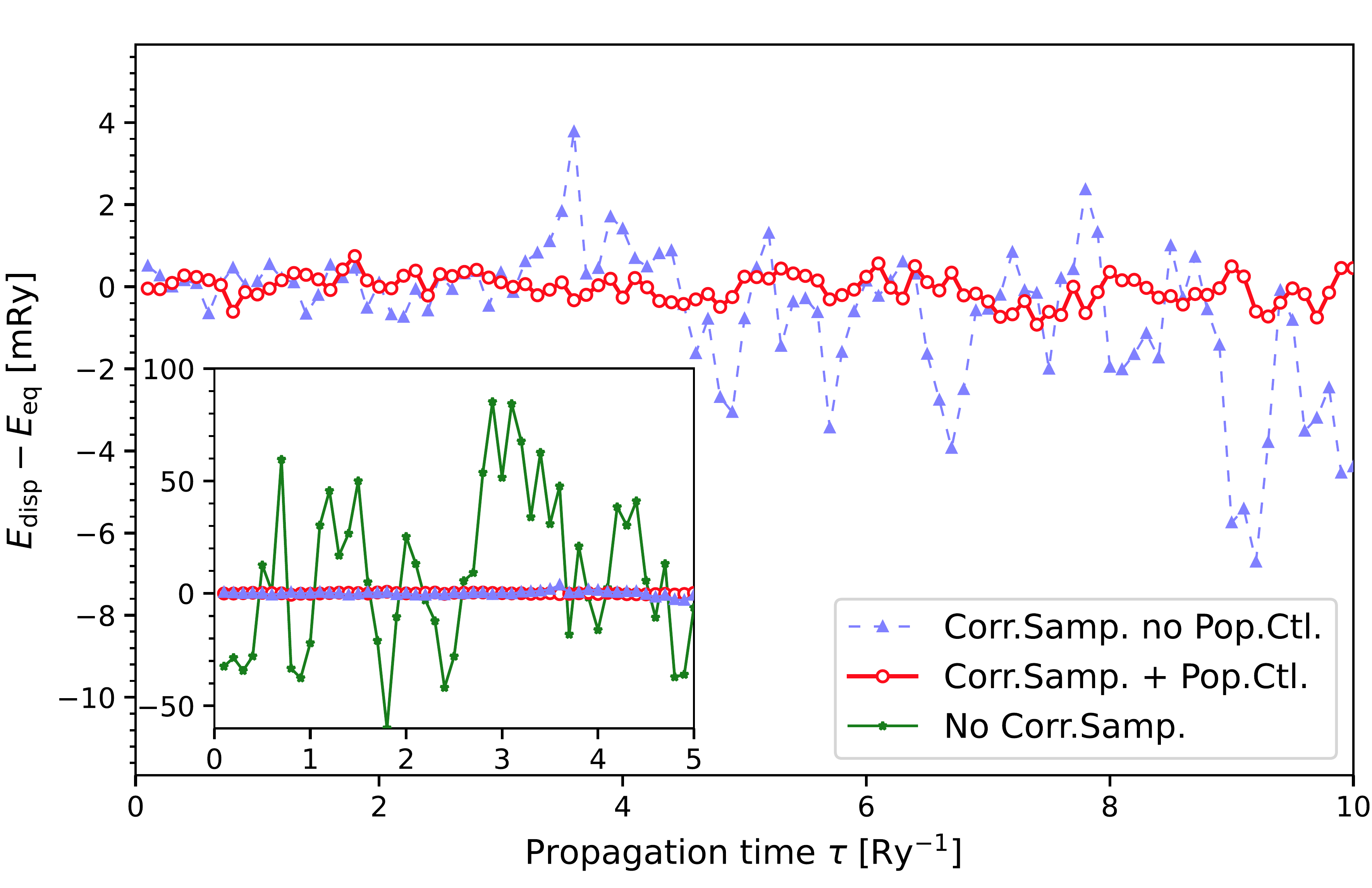}
\caption{\label{fig:Si_compare} 
Performance improvement by correlated sampling with population control in ab initio computations of the real material silicon. 
The difference in ground-state energies between two structures in bulk Si is computed [$E_\mathrm{eq}$ at equilibrium (diamond structure), and $E_\mathrm{disp}$ with one of the atoms displaced by 0.01 Angstrom] by Ph-AFQMC, with and without correlated sampling or population control. The system is a $2\times2\times2$  supercell with 32 Si atoms.
Each calculation is performed with  a total population of 4,000 random walkers.
}

\end{figure}

Here we perform a complimentary set of tests in ab initio calculations 
of real materials, in bulk silicon.  
Our computations use plane-wave basis AFQMC \cite{Suewattana_PRB_2007}
with multi-projector norm-conserving pseudopotentials \cite{Hamann_PRB_2013,Ma_MultipleProj_PW-AFQMC-PhysRevB.95.165103}.
We focus on the diamond-structured Si, with  32 atoms
in a body-centered cubic (BCC) supercell.
(We apply a twist boundary condition to the supercell using as ${\mathbf k}$-point
the BCC Baldereschi point: $(\frac{1}{4},\frac{1}{4},-\frac{1}{12})$ in fractional coordinates.)
As such, the system is an interacting many-body system with over 100 electrons 
and in excess of 9,000
plane-waves (after the use of pseudopotentials), 
presenting a stringent test of any correlated sampling approach.  
We measure the energy difference $\Delta E$ between two systems: one at the minimum-energy or equilibrium geometry, while the other with 
one of the 32 Si atoms displaced by 0.01 Angstrom. 
The energy difference between the two  
systems can be used 
to compute the force exerted on the displaced atom. Moreover,
such energy differences are crucial for structural optimizations or 
reaction pathway studies. Here, the static version of 
correlated sampling population control is tested. 

Figure~\ref{fig:Si_compare} shows a comparison of the computed energy difference of the two systems, using Ph-AFQMC without correlated sampling, with correlated sampling but without population control, and finally with our full algorithm of correlated sampling with population control. The total population size (reflecting the total computational cost) is the same for 
all three.  
The computed energy difference is around zero, since the structure is 
at equilibrium and the atomic force on the displaced atom vanishes. 
Without correlated sampling, the calculations show fluctuations which are 
 1-2 orders of magnitude larger than with correlated sampling, as seen in the inset. The main panel omits these results and show only a magnified view of 
 the two correlated sampling results. 
 Without population control, correlated sampling exhibits a clear increase in statistical fluctuations with 
 projection time. With population control, the fluctuations are reduced. 
 Furthermore, the growth with random walk steps is 
 much 
 suppressed; in fact it is barely discernible in the range of projection time studied, which is far greater than the time needed for the targeted difference to reach convergence. 

\section{Conclusion and outlook}\label{sec:conclusion}
In this work, we proposed a branching and population control algorithm for correlated sampling. The algorithm is generally applicable to Monte Carlo calculations that involve branching random walks. We outlined several variants for implementing the algorithm, 
which can be adapted based on the calculational setup and the behavior of the underlying correlated sampling method. We also discussed the quantification and validation of the effectiveness of correlated sampling. We illustrated and tested our algorithm in the Hubbard model and in ab initio solid calculations, using the ground-state AFQMC method. 
The population control algorithm was shown to significantly increase the efficiency of correlated sampling, and extend the duration of correlation as a function of random walk time. 

We expect that the algorithm will expand the range of applicability of correlated sampling, to larger systems and more diverse and challenging problems. Many applications in different areas can be pursued along these lines. In addition, there is considerable room to improve the algorithm itself. For example, in defining the function $f$ or in implementing branching/population control after the reference weights are produced, we have only taken a first step, and expect considerable efficiency gain can be available with further study.

\bibliographystyle{unsrt}
\bibliography{ref.bib}

\end{document}